\documentclass{pasj00}
\draft

\begin{document}
\SetRunningHead{Wang et al.}{Wang et al.}
\Received{2008//}
\Accepted{//}

\newcommand{\degpoint}{\mbox{$^\circ\mskip-7.0mu.\,$}}
\newcommand{\secpoint}{\mbox{$''\mskip-7.6mu.\,$}}
\newcommand{\lya}{Lyman~$\alpha\,\,$}
\newcommand{\rAA}{{\AA \enskip}}
\def\arcs{$''~$}

\title{A massive disk galaxy at $z>3$ along the sightline of QSO 1508+5714\thanks{Based on data collected at the Subaru Telescope,
which is operated by the National Astronomical Observatory of
Japan.}\thanks{We acknowledge the use of HST WFPC2 archive data}}

\author{Yiping \textsc {Wang}}
\affil{National Astronomical
observatories of China, Chinese Academy of Sciences, Beijing, China} \email{ypwang@bao.ac.cn}

\author{Toru \textsc{Yamada}}
\affil{Tohoku University, Aramaki, Aoba, Sendai 980-8578, Japan}

\author{Ichi {\sc Tanaka}}
\affil{Subaru telescope, National Astronomical Obs. of Japan} \and
\author{Masanori {\sc Iye}}
\affil{National Astronomical Obs. of Japan, Mitaka, Japan}


\maketitle

\begin{abstract}
We have obtained deep images in the $\it BVRIJHKs$ bands of the
field centered on QSO 1508+5714 ($z_{em} =4.28$) with the Suprime
camera, FOCAS and MOIRCS cameras on Subaru telescope. We report here the detection of a B-dropout galaxy, which is $3\secpoint 5$
north-west of the QSO sightline. A
photometric redshift analysis is presented to complement the color
selection. Given the photometric properties of this object
($M\,=\,-22.2$, making $\rm L\approx 3\, L^{\ast}$, if placed at its photometric redshift
$z\sim 3.5$), as well as the S$\acute{e}$rsic index ($\rm n \sim 1$)
derived from a 2-D imaging decomposition of the HST WFPC2 image taken in the $I_{\rm F814}$ filter, the
identified system is consistent with a massive disk galaxy at $z>3$. If confirmed, it would be one of the most distant massive
disk galaxies known so far.
\end{abstract}

\section{Introduction}
Based on the cold dark matter theory, the models of ``hierarchical formation'' give a step-wise 
process in which small objects merge, inducing bursts of star formation that create different 
types of massive and bright galaxies which we see today \citep{Sea78}. In contrast, the monolithic models 
suggest a top-down formation scenario assuming that stars form through a direct collapse of 
a large gas cloud, and flatten over time as the gas cools \citep{Egg62}. 

According to hierarchical galaxy formation models, galactic sizes should grow with time, and 
massive disk galaxies were formed at $z\lesssim 1$ \citep{Fal80, Mo98}. To place tight 
constraints on the galaxy formation scenario, observational studies of high redshift galaxies 
on the size, morphology, stellar mass and age et al. have been carried out over the past 
decade \citep{Vog96, Lil98, Sim99}. Studies of the galaxy size evolution at high redshift 
($z\sim 2-6$) from the Great Observatories Origins Deep Survey(GOODS), as well as Ultra 
Deep Field (UDF) and UDF-parallel Advanced Camera for Surveys fields showed a clear 
decrease in size from $z\sim 1$ to $\rm 4$ and a possible extension to $z\sim 6$, indicating 
that high redshift galaxies are compact in size ($\sim 0\secpoint 1 \,-\, 0\secpoint 3$) and that large 
($\gtsim 0\secpoint 4, \,\gtsim 3\,kpc$) low surface brightness galaxies are 
rare \citep{Bou04, Fer04}. On the other hand, observational evidences of massive disk galaxies at redshift 
much beyond one has become considerably abundant in literatures recently \citep{Liu00, Cim04, Yan04, Dad05, 
Fu05, Kri06, Pap06}. The most direct edvidences for the existence of large disks at high redshift 
are usually from the deep near-infrared imaging of high redshift QSO field, clusters and Hubble 
Deep Field-South \citep{Iye00, Dok01, Iye03, Lab03, Sto08, McG08}. This is probably because previous 
U-dropout technique revealed only the unobscured star forming regions rather than the more 
evolved underlying population that forms the disk, and multiwavelength surveys with high quality 
photometry are suggested for an unbiased census of massive galaxies in the early 
universe \citep{Gia96, Low97, Dok06}.

In this paper, we present the results from $\it BVRIJHKs$
broad-band deep images of the QSO 1508+5714 field ($z_{\rm
em}=4.28$), as well as the analysis of the rest frame UV/optical morphology and 
the SED of a B-dropout galaxy nearby QSO. In addition, the low dispersion spectrum of the QSO, showing clearly the strong and
sharp \lya  and $\rm CIV$ emission lines, as well as a Lyman Limit
System at $z=3.88$ ($\tau_{\rm LLS}>4.6$, corresponding to a
neutral hydrogen column density of $\rm >7\times
10^{17}\,cm^{-2}$) \citep{Sto94, Sto96}. A possible connection between the Lyman Limit absorption and the galaxy would be discussed in the next paper. The cosmological parameters $\Omega=0.27,\,\,\Lambda=0.73$
and $\rm H=71\,km/s/Mpc$ are adopted throughout.

\section{Observation and data reduction}
The $\it VRI$ deep imaging of the QSO1508+5714 field was made on
May 28, 2001(UT) during a test run, using the FOCAS camera on
Subaru 8.2m telescope at Mauna Kea. The camera is made of two $\rm
2048\,\,x\,\,4096$ CCDs and a pixel scale of $0\secpoint 104$,
providing a field size of $6'\times 6'$ \citep{Kas02}. We adopted in the
observation a binned mode to improve the signal-to-noise ratio,
resulting in a pixel scale of $0\secpoint 208$. The total exposure
times were 2800, 1500 and 1700 seconds for $\it V$($5500\rAA$),
$\it R$($6600\rAA$) and $\it I$($8050\rAA$) respectively.

The data reduction was performed using the IRAF package. After bias
subtraction, each frame was subsequently flat-fielded using a
combination of dome flats or twilight sky flats to remove
pixel-to-pixel variations across the CCD chips. We perform sky
background subtraction by removing a 2nd-order polynomial fitted
to the sky components. Finally, the dithered frames within
each bandpass were averaged using an outlier rejection algorithm. The weather
was clear and the observational condition was good. The final
combined image has a stellar PSF with full width at half maximum
(FWHM)$\sim 0\secpoint 7$ in the $\it R$ band.

In order to constrain the redshift range of the detected galaxies,
we obtained a one hour deep-B exposure by service observing mode on
July 14, 2007(UT) with the Subaru Suprime camera, which is a
mosaic of ten $\rm 2K\times 4K$ CCDs and covers a $34' \times 27'$
field of view with a pixel scale of $0\secpoint 20$ \citep{Miy02}. The final
combined image has a stellar PSF with full width at half maximum
(FWHM)$\sim 0\secpoint 7$. We also
obtained 450, 1260 and 750 seconds exposure in the $\it
J$($1.26\mu m$), $\it H$($1.64\mu m$) and $\it Ks$($2.14\mu m$)
bands on June 8 and June 25, 2007(UT) using the Subaru MOIRCS
camera, which is a wide-field imaging camera and spectrograph,
with a field view of $4' \times 7'$ and a spatial resolution of
$0\secpoint 117$/pixel \citep{ich06,suz08}. We use data reduction packages ``SDFRED''
for the Suprime-Cam image \citep{Yag02,Ouc04}, as well as the package "MCSRED"
developed by Ichi Tanaka (2008) for the MOIRCS data, to reduce the
raw data and produce the scientific images for further photometry. The weather
was clear and we have gotten a stellar PSF with full width at half maximum
(FWHM)$\sim 0\secpoint 3$ in the $\it Ks$ band for the final
combined image.

We retrieved from the HST data archive a 4800 seconds exposure
taken by the Hubble Space Telescope (HST) WFPC2 in the F814W-band
(approximately $\it I$ band, and a pixel scale of $0\secpoint 0996$). Considering the high resolution of
the HST image, we will rely on it for the morphological analysis
of the galaxy candidates.

For the current study, we have performed: i) careful PSF
subtraction on the combined images of all bandpasses, to reduce
the effects of QSO light. The modeled PSF was determined using a
set of bright stars in the same image with DAOPHOT; ii) Because
our image is more sensitive in the $\it R$ than in the $\it I$
band, the object detection was done by running the SExtractor package
on the combined and PSF subtracted image in the $\it R$-band, and
a detection threshold of $\mu =3\sigma$ of the skylevel was
adopted (Bertin \& Arnouts 1996).  Colors were determined by
re-running SExtractor in the double-image mode, in which the faint
objects detected on the ``detection image''(in the $\it R$-band)
were measured with the same aperture in the registered other
bands. The photometric results are shown in Tab.\ref{tab:color1}
and Tab.\ref{tab:color2}, where the data with footnote ${a}$ for
galaxy "G1" and "G2" are measured with a small diameter aperture
($1\secpoint 25$) using SExtractor and those with footnote ${b}$
are measured with a large diameter aperture ($2\secpoint 5$) using
SExtractor.

Also observed were several standard stars selected from Landolt
(1992) for $\it BVRI$ bands on the same night and at similar
airmasses. The standard star FS27 was observed as the photometric
calibrator for $\it JHKs$ bands, which was selected from Hunt et
al. (1998) and the 2MASS All-Sky catalogue of Point Sources. We have
adopted the equation B(AB)=B-0.11, V(AB)=V+0.02, R(AB)=R+0.20,
I(AB)=I+0.45, J(AB)=J+0.9, H(AB)=H+1.38, Ks(AB)=Ks+1.86, to put
the magnitude onto the AB system \citep{Fuk95, Bes88}. All magnitudes subsequently quoted in this paper are
on the AB system.

\section{Analysis}
\subsection{Galaxies near the QSO sightline and their colors }

\begin{figure}\begin{center}
\FigureFile(90mm,90mm){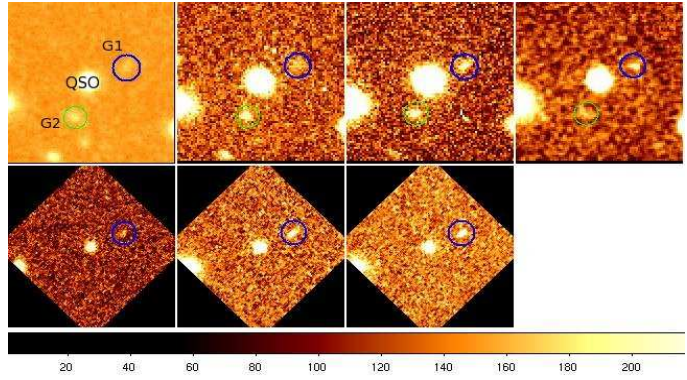}
\end{center}
\caption{ The combined images of
a region of $\sim 15\times 15$ $\rm arcsec^2$ nearby the QSO in the $\it BVRIJHKs$
bands (top figures: B,V,R,I bands from left to right, and bottom figures: J, H, Ks bands from left to right). Galaxy
``G1'' is indicated by blue circles, which is located in the northwest of the QSO sightline in the image, and "G2'' is marked by green circles in the southeast of the QSO. The diameters of circles match roughly the aperture diameters for the photometry, i.e. $\sim 2\secpoint 5$. North is up and East
to the left.}\label{fig:15}
\end{figure}

We show in Fig.\ref{fig:15} the combined images of $\sim$ 15\arcs square region 
surrounding the QSO in the $\it BVRIJHKs$ bands. Two objects standout in the images
of $\it BVRI$ bands, within a
distance of $3\secpoint 5$ northwest and southeast of the QSO line
of sight, which are designated as ``G1'' and ``G2'' in this work. 
Galaxy ``G1'' is marked by blue circles in the images, and ``G2'' is marked
by green circles in the images of $\it BVRI$ bands. Galaxy ``G2'' is not detected in the NIR 
images.

Firstly, we adopt the V-dropout selection criteria
designed by Fukugita et al.(2004) to select galaxies at $z \sim 4$, 
who use the same set of filters
as ours to select galaxies at $z\sim 4$. 
\begin{eqnarray}
0.95<V-R<2.0,\cr ~~~~~~~~~~0.59(R-I)+0.54< V-R< 3.6(R-I)+0.4,\cr
23.5<I,~~~R-I<1.0.~~~~ \label{eq:colcond}
\end{eqnarray}

The colors of galaxies
"G1" and "G2", with small diameter aperture ($1\secpoint 25$) and
large diameter aperture ($2\secpoint 5$), are shown in
Tab.\ref{tab:color2}. Considering the peculiar colors of galaxy ``G1'' with $V-R > 1.0, R-I < 0.7$ , we tentatively
suggest that it is most likely a candidate galaxy at redshift $z\sim 4$, and might be responsible for the
Lyman Limit absorption at $z=3.88$ seen in the QSO spectrum.

However, we understand that the $\it VRI$ filter set
is not ideal for two-color photometric selection of galaxies at
$z\sim 4$, according to various model calculations \citep{Ste01}.
This is because an evolved galaxy at $z\sim 0.5$ is likely to have
similar colors to the high redshift object owing to the presence
of the $4000\rAA$ break. We therefore made deep B-imaging of the
field centered on QSO1508+5714 with the
Suprime camera on Subaru telescope to constrain the redshift range
of galaxies ``G1'' and ``G2''.

With one hour exposure, we clearly detected galaxy ``G2'', with a
detection threshold of $>3\sigma$ over the sky level. However,
galaxy ``G1'' is much fainter, and gives a $2\sigma$ detection of
$\rm B=27.92$. This means that the spectrum properties of galaxy
``G1'' do show the Lyman break, and indicates that galaxy ``G1''
might be a high redshift galaxy at $z\sim 4$. Comparing with the
BRI color selection for $z\sim 4$ LBGs given by Prochaska et al.
(2002), we found that the colors of galaxy ``G1'' satisfy their
criteria, but not for galaxy ``G2''.

\begin{eqnarray}
B-R>2.0,\cr ~~~~~~~~~~B-R>1.2\,(R-I)+1.6,\cr ~~~~~~~R-I<1.0.
\label{eq:colcond}
\end{eqnarray}

Although we cannot completely exclude the possibility
that galaxy ``G2'' is also at high-z with $ B-R=1.73\pm 0.07$, we
suspect that it is most likely a foreground contaminator. The reasons are as following: 1)considering the surface density of Lyman Break
Galaxies at $z\sim 4$, down to magnitude $\rm R=25.5$ ($\sim 1$
per square arcmin), the probability of finding two objects at such
high redshift and within such a small area ($\sim$ 15\arcs square)
is very low; 2)meanwhile, inferred from the presence of only one
absorption line system in the QSO spectrum \citep{Ste99}, we suspect that most likely only galaxy ``G1'' is a high-z object.

\begin{longtable}{lllll}
\caption{Photometric results 1. The footnotes {a} and {b} for "G1"
and "G2" indicate the measurements with small and large apertures
described in detail in Sec. 2.}\label{tab:color1}
 \hline\hline
 $\rm Obj $ & $\rm B_{_{AB}}$ & $\rm V_{_{AB}}$ & $\rm R_{_{AB}}$ & $\rm I_{_{AB}}$ \\
 \endhead
 \hline
 \endfoot
 \hline
 \endlastfoot
 \hline
$G1_{a}$ & $27.92$ $\pm$ 0.08  & $ 27.06$ $\pm$ 0.15 & 25.68 $\pm$ 0.06 & $ 25.01$ $\pm$ 0.10 \\
$G2_{a}$ & 27.25 $\pm$ 0.05    & $ 26.34$ $\pm$ 0.08 & 25.53 $\pm$ 0.06 & $ 25.50$ $\pm$ 0.16 \\
$G1_{b}$ & $27.53$ $\pm$ 0.10  & $ 26.09$ $\pm$ 0.12 & 24.92 $\pm$ 0.06 & $ 24.39$ $\pm$ 0.11 \\
$G2_{b}$ & 26.51 $\pm$ 0.04    & $ 25.62$ $\pm$ 0.08 & 24.78 $\pm$ 0.06 & $ 24.31$ $\pm$ 0.11 \\
\hline\hline
\end{longtable}

\begin{longtable}{lllllll}
\caption{Photometric results 2. Galaxy "G2" is not detected by the
JHKs deep imaging.}\label{tab:color2}
 \hline\hline
 $\rm Obj $ & $\rm (B-R)_{AB}$ & $\rm (V-R)_{AB}$ & $\rm (R-I)_{AB}$ & $\rm J_{_{AB}}$ & $\rm H_{_{AB}}$ & $\rm Ks_{_{AB}}$ \\
 \endfirsthead
 \endfoot
 \hline
 \endlastfoot
 \hline
 $G1_{a}$ & 2.24 $\pm$ 0.1    & 1.38 $\pm$ 0.16  & 0.67 $\pm$ 0.12 & $ 22.98$ $\pm$ 0.18 & 22.60 $\pm$ 0.10 & 21.87 $\pm$ 0.08 \\
 $G2_{a}$ & 1.72 $\pm$ 0.08   & 0.81 $\pm$ 0.1   & 0.03 $\pm$ 0.17 &           $ - $     &       $ - $      &   $ - $                                                      \\
 $G1_{b}$ & 2.61 $\pm$ 0.12   & 1.17 $\pm$ 0.13  & 0.53 $\pm$ 0.12 & $ 23.00$ $\pm$ 0.27 & 22.37 $\pm$ 0.10 & 21.71 $\pm$ 0.09 \\
 $G2_{b}$ & 1.73 $\pm$ 0.07   & 0.84 $\pm$ 0.1   & 0.47 $\pm$ 0.12 &           $ - $     &   $ - $          &  $ - $           \\
 \hline\hline
\end{longtable}

\subsection{Photometric redshifts}
We used the public code (Hyperz) to obtain the photometric
redshift for the galaxy candidate, $z_{\rm phot}$. This code
uses SED fitting through a standard $\chi^2$ minimization
procedure.
The photometric uncertainties in the $\it BVRIJHKs$ fluxes of the
object are accounted for and the fluxes are compared with a set of
template spectra. For a complete description of the code and the
accuracy of its results, we refer the readers to Bolzonella et
al.(2000).

We adopted a full range of solar metallicity stellar population
models, including models matching the sequence of colors from
E-S0 to Sd, as well as a single starburst model, which was built
by the Bruzual \& Charlot evolutionary code (BC03, Bruzual \&
Charlot 2003). The template SEDs were reddened by applying the
Calzetti reddening law with a wide range of reddening values from
$\rm E_{B-V}=0.0$ to $\rm E_{B-V}=2$, with steps of $0.2$
\citep{Cal00}. The fitting procedure allowed redshifts in the range
of $0\le z \le 6$.

\begin{figure}\begin{center}
\FigureFile(90mm,90mm){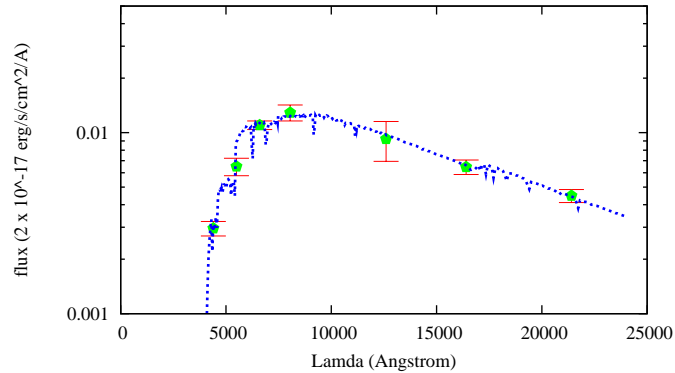}
\end{center}
\caption{Broad-band photometry of
galaxy "G1" in the $\it BVRIJHKs$ bands. The best-fit evolutionary
synthesis model is a young starburst of age of $\sim 2$ Myr at
$z\sim 3.5$ with extinction of A$_{\rm V}\sim 2$.}\label{sed}
\end{figure}

Fig. \ref{sed} shows the best fit SED for galaxy "G1" from our multi-band
photometry, which is a $\sim 2$ Myr old young starburst at $z\sim 3.5$ with A$_{\rm V}\sim 2$mag,
giving a fitting result of $\chi^2\sim 0.9$ and the corresponding
probablity of $49\%$. The estimated error on the redshift is 0.05(0.11) at
$90\%$($99\%$) confidence level. The NIR color of galaxy "G1" ($J-Ks=1.29\pm
0.28$ and $H-Ks=0.66\pm 0.13$) indicates that galaxy "G1" is
slightly bluer than the distant red galaxies (DRGs) at $2<z<4.5$,
which would have a strong Balmer/4000\rAA break. The stellar mass of galaxy ``G1'' is estimated
from the multi-band photometry SED fitting. We get a value of $\sim 8.9\times 10^{10}\,M_{\odot}$, 
which is consistent with the
relation between stellar mass and observed total $Ks$ magnitude
for galaxies at $2<z<3$ in the FIRES, GOODS and MUSYC fields, given by van Dokkum et al. (2006).

Galaxy "G2" is not detected in the $\it JHKs$ bands. We will not include its fitting
result here, due to the poor accuracy of the photometric redshift
calculation. Even for galaxy "G1", we understand that the low detection limits in $\it BVJ$ bands ($S/N < 3$) would affect the precision of the photometric redshift estimation. This is because it would cause relatively "flat" probability function due to a lack of sufficient photometric information. For detailed scientific research, further spectroscopy to
confirm the redshift of the candidates is strongly required.

\subsection{Size and morphology}

\begin{figure}\begin{center}
\FigureFile(90mm,90mm){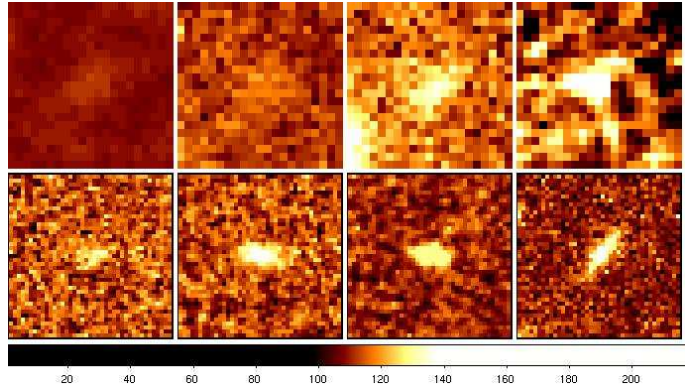}
\end{center}
\caption{B, V, R, I, J, H, Ks, and WFPC2/F814 images of galaxy ``G1''(top figures: B,V,R,I bands from left to right, and bottom figures: J, H, Ks, and WFPC2/F814 bands from left to right). The image size is $\sim 4.2\times 4.2$ $\rm arcsec^2$.
For the orientation of images in $\it BVRI$ bands, north is up and 
east to the left. North is 20\degpoint 77 left of upper and east is left of it for the WFPC2/F814 filter. For the MOIRCS images in JHKs bands, the orientation is different, with north is 40\degpoint 0 right of up and east is left of it. We show images here in its original orientation to avoid distortion.}\label{montage}
\end{figure}

To study the morphologies and sizes of both detected objects, we
estimated their structural parameters by running a 2-D fitting
algorithm GALFIT \citep{Pen02} on the HST WFPC2 archive image in the F814W
filter. We also show a montage figures of galaxy ``G1''
in Fig.\ref{montage}, to confirm the disk morphology in other bands.

\begin{figure}\begin{center}
\FigureFile(90mm,90mm){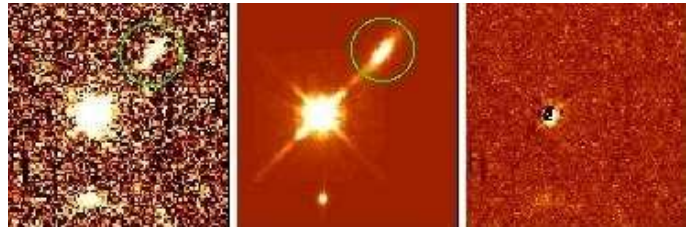}
\end{center}
\caption{2-D image decomposition on the HST WFPC2+F814W data of
the QSO field ($\sim$ 10\arcs $\times$ 10\arcs), centered on QSO 1508+5714.
From left to right, they are the original image, the final model
of the objects and the residuals from GALFIT. North is 20\degpoint 77 left of upper and east is left
of it.}\label{fig:block}
\end{figure}

\begin{figure}\begin{center}
\FigureFile(90mm,90mm){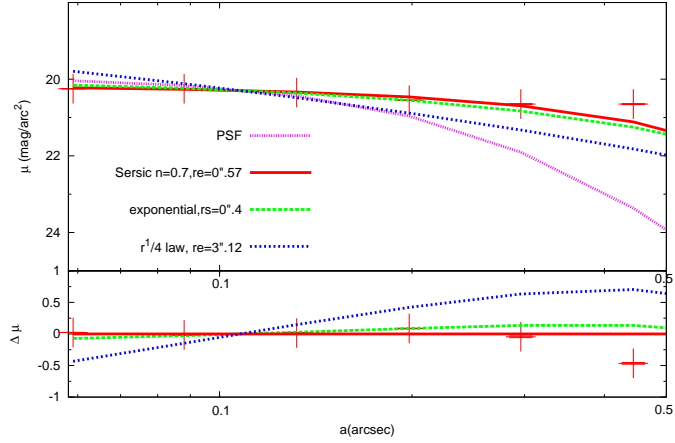}
\end{center}
\caption{Radial surface brightness profile of the WFPC2 F814W image of
the galaxy ``G1'', with best-fit S$\acute{e}$rsic, exponential, $\rm r^{1/4}$ profiles 
shown. The profiles are shown at the top panel and the deviations of the 
observed profile and two other models from the best-fit Sersic profile are 
given at the bottom.}\label{fig:prf.g1}
\end{figure}

We used the public software Tiny-Tim to create a PSF for the
convolution of the HST WFPC2 image, and made a model fitting to
the QSO field where the QSO (PSF) and the two galaxies ``G1'' and
``G2''(S$\acute{e}$rsic profiles) are fitted simultaneously to
deblend everything together, and to reduce the contaminating flux
from the wings in the PSF of the QSO. Haeussler et al.(2007) shows
that this kind of simultaneous fitting gives the most reliable
results against the neighboring contamination, especially in the analysis of deeper cosmological images or of more crowded fields. This is because the simultaneous fit of the profiles of multiple companions thereby deblends their effect on the fit to the galaxy of interest. The output images from GALFIT are presented in Fig.
\ref{fig:block}, which shows from left to right, the original
image specified by the convolution box size, the final model of
the objects in the selected field and the residual image by
subtracting the second from the first image. The surface brightness radial 
profiles of galaxy ``G1'' is shown in Fig.\ref{fig:prf.g1}, which is measured by fitting ellipses to the WFPC2 images with the STSDAS task ELLIPSE.

For galaxy "G1", the best fit structural parameters are $\chi^2 =
1.58$, the S$\acute{e}$rsic index $n=0.7\pm 0.2$, and an
effective radius $r_{e}=0\secpoint 57\pm 0\secpoint 23$. In
addition, we have done several checks of the systematics, such as,
by masking out the central region of the QSO which cannot be
fitted well, and running again the simultaneous model fitting to
the QSO field. We found that there is no significant systematic
errors for the current results, especially for the S$\acute{e}$rsic index.

However, the WFPC2/F814 image maps the unobscured star-forming regions at rest
frame UV wavelengths, it would not be appropriate for discussing the radial profile or morphology of the galaxy. On the other hand, MOIRCS Ks-band imaging is looking at the rest-optical
wavelength which has a stellar PSF with full width at half maximum (FWHM) $\sim 0\secpoint 3$ for the combined image. We understand that the parameters of the S$\acute{e}$rsic profiles would be affected by seeing. In case of a $\sim 0\secpoint 3$ (FWHM) resolution, we hope to see at least the radial profile in the outer region. So, the seeing-limited
ground-based Ks imaging would be a good complement to the WFPC2/F814 results for the morphological studies.

\begin{figure}\begin{center}
\FigureFile(90mm,90mm){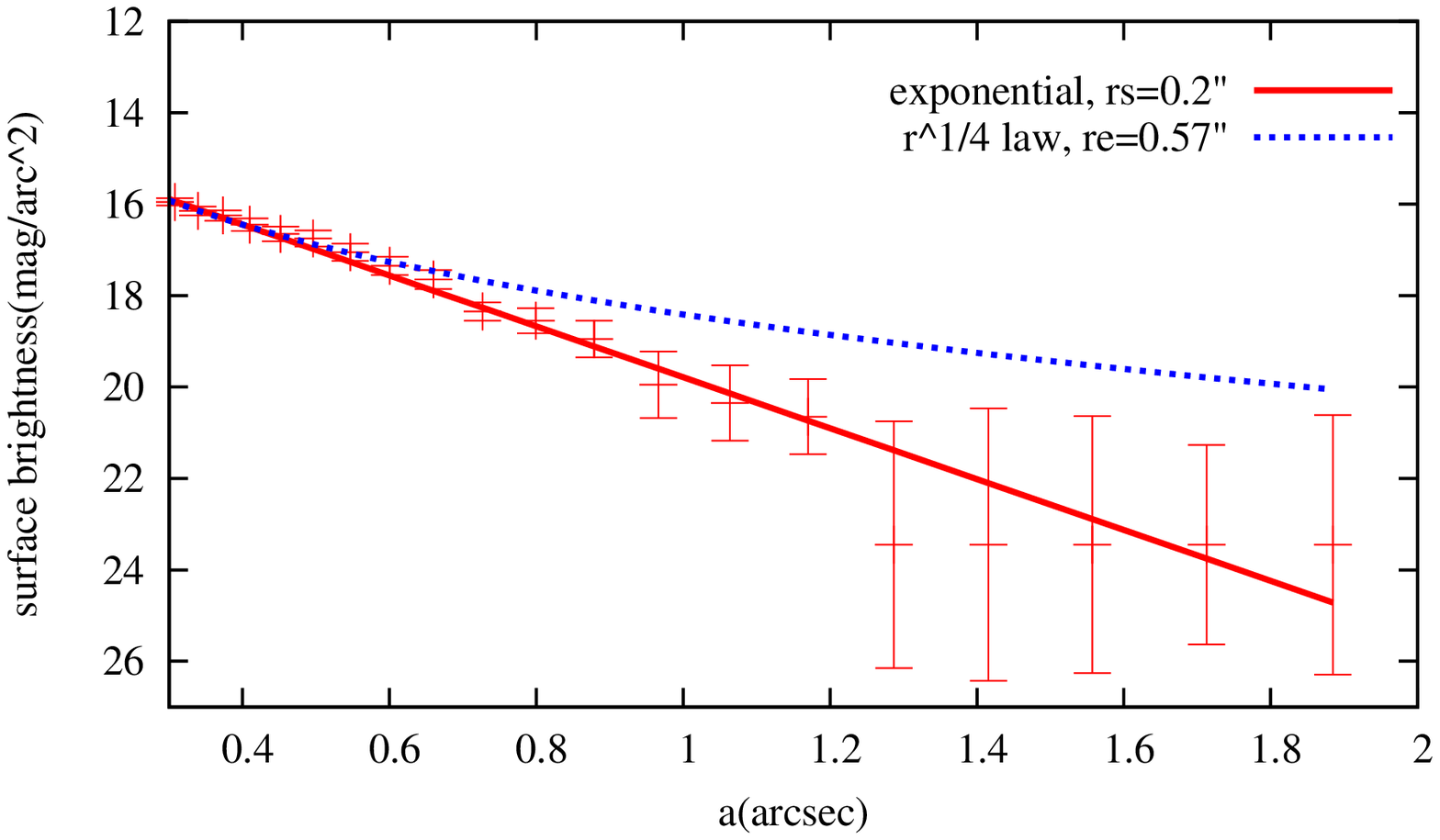}
\end{center}
\caption{Radial surface brightness profile of the MOIRCS Ks-band image of
the galaxy ``G1''. The overplotted lines are the best-fit exponential and $r^{1/4}$ laws.}\label{fig:prf.k.g1}
\end{figure}

Similar as we have done for the WFPC2 F814 image, we made a GALFIT fitting to the Ks image by convolving a PSF to the models. The PSF image for convolution is created by fitting a nearby bright star with a number of S$\acute{e}$rsic profiles. 
The surface brightness 
radial profile of galaxy ``G1'' is shown in Fig.\ref{fig:prf.k.g1}, which is measured by fitting ellipses to the WFPC2 images with the STSDAS task ELLIPSE. The overplotted lines are the best-fit exponential and $r^{1/4}$ laws from GALFIT fitting.

The measured angular size corresponds to a physical radii of $\rm
r_{e}\sim 3\,kpc$ at $z\sim 3.5$. According to the relationship between the S$\acute{e}$rsic index
($\rm n$) and the morphological type, we propose that galaxy
``G1'' is most probably a large disk galaxy at high-z \citep{deJ96, Rav04}.

The studies of the size and
morphology of distant objects have been carried out by several
authors, and the detections of large disk galaxies at high
redshift have been accumulating \citep{Iye03, Lab03, Bou04, Fer04,
Ove08}. Recently, a few large disk systems at $z\sim 2.5 - 3$,
with $r_{e}\sim 3.5 - 7\,kpc$ are reported by Stockton et al.
(2008) and Akiyama et al. (2008). If the redshift of galaxy "G1"
could be confirmed, it would give evidence for the existence of
such large disks at even higher redshift, which might be the
progenitors of the similar systems detected by Stockton et al.
(2008) and Akiyama et al. (2008). Considering that most of the large disks detected
by other groups are old galaxies, galaxy ``G1'' has a bluer color, and a young ``dusty'' starburst spectrum. This provides strong evidence for a large disk formation in 
the early universe with on-going star formation. Meanwhile, the presence of such
a population at high redshift would require the current galaxy
formation models to allow for the presence of early-forming
massive disks.

\section{Summary}
We have presented an analysis of deep images in the $\it BVRIJHKs$ bands centered on QSO 1508+5714 at $z_{em}=4.28$, which
is known to contain a Lyman Limit System at $z_{abs}=3.88$ seen in
the QSO spectrum.

A B-dropout galaxy ``G1'' which is about $3\secpoint 5$ northwest of the QSO
sightline is clearly detected. To complement the color
selection, a photometric redshift analysis is presented and gives a value of $z_{\rm phot}\sim 3.5$. We estimate its projected distance of $\rm
24\,kpc$ at $z \sim 3.5$, and further derive the absolute magnitude $\rm M=
-22.2$, making it about $\sim 3\,L^{\ast}$ for a star-forming
galaxy spectrum. The stellar mass of galaxy ``G1'' is estimated from the multi-band photometry SED fitting
method. We get a value of $\sim 8.9\times 10^{10} M_{\odot}$, consistent with the
relation between stellar mass and observed total $Ks$ magnitude
for galaxies at $2<z<3$ in the FIRES, GOODS and MUSYC fields, given by van Dokkum et al. (2006).

We have run a 2-D imaging decomposition on the HST WFPC2 archive
image of the QSO nearby field by GALFIT, which gives a
S$\acute{e}$rsic index ($\rm n\sim 1$) for galaxy ``G1'',
indicating a possible late-type morphology according to the
S$\acute{e}$rsic index--morphological type relation. The radial profile of the seeing-limited ground-based Ks image is shown as a 
complement to the WFPC2/F814 results (rest frame UV wavelengths) for the morphological studies, since it is looking at the rest-optical wavelength. We can see in Fig.6 the radial profile can be well described by exponential curve at least at the larger radii ($>0\secpoint 5$).
If galaxy
``G1'' is placed at $z \sim 3.5$, we derive its effective
radius ($\rm r_{e}\sim 3\,kpc$), with the WMAP cosmology adopted by
this study. 

Given the small impact parameter, the peculiar colors, the
estimation of the photometric redshift, as well as various lines
of evidence discussed in the previous sections, we suspect that
galaxy ``G1'' is most likely to be a massive disk at high redshift,
and might give rise to the known LLS ($z_{abs}=3.88$) seen in the QSO
spectrum.

Such a large disk galaxy at high redshift would have a strong impact on the
galaxy formation theory, if the redshift of galaxy "G1" could be
confirmed. Thus, a spectroscopic redshift is needed to test these conclusions.

\section*{acknowledgments}
This work is supported by National Scientific Fundation of China
(NSFC 10173025, 10673013 and 10778709) and the Chinese 973 project(TG 2000077602). YPW
acknowledges the Subaru team for the hospitality. YPW would thank
Dr. Peng for his kindly help with the GALFIT code, Dr. Furusawa
for the Suprime-Cam observation and data reduction. We thank the anonymous referee for the helpful comments and suggestions.


\end{document}